\documentclass[a4paper,twoside,10pt]{article}
\usepackage{graphicx,publaob}
\usepackage{wrapfig}
\def\papertype{Invited}
\begin{document}

% Title
\title{SPECTROSCOPY AND SPECTROPOLARIMETRY OF AGN: FROM OBSERVATIONS TO MODELLING}

% Authors
\authors{L. \v C. POPOVI\'C$^{1,2,3}$, D. ILI\'C$^1$,  E. BON$^2$, N. BON$^2$, P. JOVANOVI\'C$^2$, A. KOVA\v CEVI\'C$^1$,}
\authors{J. KOVA\v CEVI\'c -DOJ\v CINOVI\'C$^2$, M. LAKI\'CEVI\'C$^2$, S. MAR\v CETA-MANDI\'C$^1$,}
\authors{N. RAKI\'C$^3$, D\lowercase{j}. SAVI\'C$^2$, S. SIMI\'C$^4$ \lowercase{and} M. STALEVSKI$^2$}

% Addresses and e-mails
\address{$^1$Department of Astronomy, Faculty of Mathematics, University of Belgrade, Studentski trg 16, 11000 Belgrade, Serbia}
\Email{dilic}{matf.bg.ac}{rs}
\address{$^2$Astronomical Observatory, Volgina 7, 11000 Belgrade, Serbia}
%\Email{lpopovic}{aob}{rs}
\address{$^3$Faculty of Science, University of Banja Luka, Republic of Srpska, B\&H}
\address{$^4$Faculty of Science, University of Kragujevac, Radoja Domanovi\'ca 12, 34000 Kragujevac, Serbia}
%\Email{ssimic}{kg.ac}{rs}

% Running titles
\markboth{SPECTROSCOPY AND SPECTROPOLARIMETRY OF AGN}{L. \v C. POPOVI\'C et al.}

% Abstract
\abstract{Active galactic nuclei (AGN) are one of the most luminous objects in the
Universe, emitting powerful continuum and line emission across all wavelength
bands. They represent an important link in the investigations of the galaxy
evolution and cosmology. The resolving of the AGN inner structure is still a
difficult task with current instruments, therefore the spectroscopy and
spectropolarimetry are crucial tools to investigate these objects and their
components, such as the properties of the supermassive black hole, the broad
line region, and the dusty torus. In this review, we present the
results of the project ''Astrophysical spectroscopy of extragalactic objects'',
from the observations, data processing and analysis, to the modelling of
different regions in AGN.}

% Section and subsection
\section{INTRODUCTION}

Active galactic nuclei (AGN) are in the focus of the modern astrophysical investigation, 
since it is widely believed that all galaxies had at least one phase of high activity  (AGN phase) in their life-time. Moreover, the AGN feed back huge amount of energy into 
surrounding medium, which may have influence on all scales, from the host galaxy to the intergalactic
medium. Therefore, it is important to understand the structure of AGN and their radiation processes.

AGN host in their center super-massive black hole (SMBH) which is actively fueled by gas through 
the accretion disk. The accretion disk is emitting the X-ray continuum (and also Fe K$\alpha$ line) 
that is a powerful source of radiation which ionize the surrounding gas.
The ionized gas that is very close to the central SMBH emits broad emission lines (with full width at half 
maximum - FWHM of $>1000$ km s$^{-1}$), that is called the broad line region (BLR), whereas the ionized gas far from the center emits narrow lines (FWHM  $<1000$ km s$^{-1}$), and consequently it is called the narrow line region (NLR). The BLR  sometimes can be obscured by the dusty torus-like region, depending on 
the orientation of the system. Using the obscuration as a criterion, we can divide AGN into two classes: 
the type 1 AGN (the BLR is not obscured), which show the broad and narrow
emission lines in the UV/optical/IR spectra, and
type 2 (the BLR is obscured) AGN, which have only narrow emission lines. Even though we know the general model
of these objects, there are still many open questions. Some important ones are: i) what is the 
structure and kinematics of the BLR; ii) how to estimate 
the mass of the central SMBH, iii) are there present and how to detect binary SMBH, 
iv) what is the structure of the dusty torus, and many others. In order to answer 
these, the spectroscopic and spectropolarimetric observations can give us great insight into these 
hidden regions. 

The investigation the AGN structure using spectroscopy, spectropolarimetry and other methods/effects (simulations,
gravitational milli- and micro-lensing, etc.) have been a subject of the project 
''Astrophysical spectroscopy of extragalactic objects'' (P.I. L.\v C. Popovi\'c),
that was accepted in 2010, and has been 
funded (until the end of 2017) by the 
Ministry of Education, Science and Technological Development
 of Republic of Serbia. In our research we try to fix some questions given above and here 
we give an overview of recently obtained results.

\section{LONG-TERM OPTICAL MONITORING OF AN AGN SAMPLE}

AGN show high variability in their spectra, that can be used 
to probe the kinematics and physics of the BLR  by comparing the
variability of the continuum and broad emission line fluxes. Therefore we performed 
the long-term optical monitoring campaign of several type 1 AGN, whose broad emission 
lines have different spectral characteristics: Seyfert 1 galaxies (NGC 5548, NGC 4151, 
NGC 7469), Narrow-line Seyfert 1 galaxy - NLSy 1 (Ark 564), double-peaked line radio loud (3C 390.3)
and radio quiet (Arp 102B) galaxy, and a luminous quasar (E1821+643).
Additionally, we explore variability of two AGN in spectro-polarization: 3C390.3 and Mrk 6.
The spectral observations were
done with six telescopes based at four different observatories: the Special Astrophysical Observatory (SAO) 
of the Russian Academy of Science in Russia (1-m and 6-m telescopes), the Guillermo Haro Astrophysical
Observatory in Mexico (2.1-m telescope), the Observatorio Astronomico Nacional at San Pedro Martir in 
Mexico (2.1-m telescope), and the Calar Alto Observatory in Spain (3.5-m and 2.2-m telescopes). 
The spectro-polarimetric observations were performed with 6 m telescope of  the 
SAO using the modified spectrograph SCORPIO.

To study AGN, we use reverberation mapping  that uses  temporal fluctuations of the central continuum
source, and the subsequent response of the BLR emission. The time delay between the continuum
and the broad line fluctuations provides an estimate of
the size of the BLR, and can also be used to estimate the
black hole mass.
Our reverberation mapping measurements of the radius of the BLR  are based  on  the 
Z-transformed Discrete Correlation Function and procedures that model the statistically likely behavior of the 
light curves in the gaps between observations (e. g. JAVELIN, Gaussian process regression-GP). 
These procedures  were applied on the continuum and line flux light-curves of our objects.
We used either observed or simulated light-curves  to get the most reliable result 
(see Kova\v cevi\'c et al. 2014b, 2015). Our reverberation measurements are included in the AGN Black Hole Mass Database (http://www.astro.gsu.edu/AGNmass/) hosted at Georgia State University, USA. It contains all AGN with published spectroscopic reverberation mapping results in the refereed journals.

\begin{wrapfigure}{r}{0.5\textwidth}
  \begin{center}
    \includegraphics[width=0.5\textwidth]{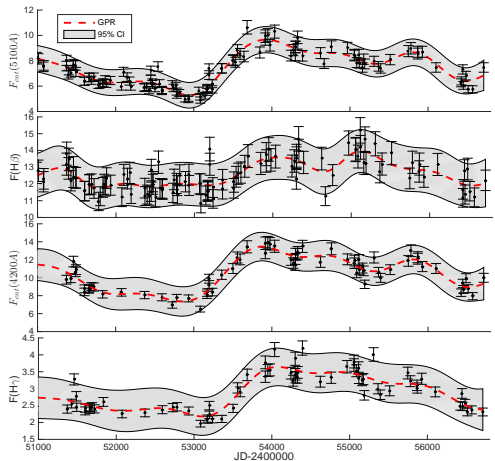}
  \end{center}
\caption{GP best fit (dashed line) to the observed light curves (dots with error bars) of the continuum at 5100\, \AA, H$\beta$, continuum at 4200\, \AA\, and H$\gamma$ (from top to bottom) in quasar E1821+643.
The shaded band represents the 95$\%$ confidence interval (CI) for the GP predicted curve (Shapovalova et al. 2016).}
\end{wrapfigure} 
Any tool describing the AGN variability has to handle irregular sampling and measurement errors in observed light curves, in order to produce  physical results. Since AGN light curves are too sparsely sampled to resolve day variability, using simple linear interpolation
between data points are impossible. Linear interpolation also incorrectly assumes that there is no uncertainty associated with the interpolation process or the measurements. For these
reasons, we model the AGN continuum light curve using a
stochastic model of AGN variability, such as Gaussian process (GP) regression, allowing us to evaluate
the light curve at arbitrarily small timescales.
The ability of GP is demonstrated in Fig. 1 (see Shapovalova et al. 2016 for details) where the  flares become clearer in the GP light curves  then in the observed curves alone. We also introduced GP for determination of periodicity in AGN light curves (Kova{\v c}evi{\'c} et  al. 2017).

The spectral data have been presented and analyzed in Shapovalova et al. (2001, 2004, 2008, 2010a,
2010b, 2012, 2013, 2016, 2017), Popovi\'c et al. (2008, 2011, 2014) and Afanasiev (2014, 2015).
Some of the important results in this part are: i)
published online spectral data (line and continuum fluxes obtained using uniform procedures) 
from several decades of monitoring for seven type 1 AGN; ii) estimated the size of the BLR and
the mass of the SMBH; iii) the BLR is probably of a disk-like shape, but with complex structure,
in the sense that the single geometry cannot explain the whole BLR (e.g. outflows or hot-spots 
are present, etc.). Also, the BLR is mainly heated by the photoionization from the central source,
but other mechanisms may be present, which is seen in the lack of correlation between line and 
continuum fluxes; iv) it seems that polarization region in AGN is smaller than we expected.
A short review of these investigations and a comparative analysis of the results are given in Ili\'c et al. (2015, 2017).

One important part of these investigations was the analysis of the long-term light curves of different
emission lines and continuum searching for periodicities (Bon et al. 2012, 2016, Kova\v cevi\'c et al. 2017).

\section{SUPER-MASSIVE BINARY BLACK HOLES}

Long term observations can indicate some type of variability in the continuum and line flux which 
has quasi-periodical (see Shapovalova et al. 2010b), 
or periodical behavior (see Bon et al. 2012, 2016, Shapovalova et al. 2016, 
Kova\v cevi\'c et al. 2017). This periodicity can be caused by 
perturbations in the emission line region (see e.g. Jovanovi\'c et al. 2010), 
but also may indicate a presence of the SMBH binary system
in the center of an AGN  (for a review see Popovi\'c 2012).  

The binary SMBH are expected to be in the center of some galaxies (Begerman et al. 1980, 
Gaskell 1983), and since they are a result of galactic mergers, they are probably
surrounded by gas, therefore one can expect that one or both black holes in the system are accreting 
matter producing radiation similar to the AGN emission (Popovi\'c 2012). 
The observations of the binary SMBH is possible on kpc-scale (see e.g. Woo 2014), however
on the distances between components of  order smaller than pc it is not possible to resolve the components by
available telescopes. Therefore the spectroscopy is the only way to detect the SMBH binary candidates.

Our investigations of the sub-pc SMBH binaries using spectroscopy have 
been performed in two directions: i) modeling the  sub-pc SMBH binary systems
in order to detect specific feature in the spectral lines and their shifts 
(Popovi\'c et al. 2000, Jovanovi\'c et al. 2014, Smailagi\'c \& Bon 2015, 
Simi\'c \& Popovi\'c 2016), and 
ii) exploring the long-term variability in the line shapes (Bon et al. 2012, 2016, Shapovalova et al. 2016;
Kova\v cevi\'c et al. 2017). Note here that we made a
discovery of the first spectroscopically resolved sub-parsec orbit of a 
SMBH binary (see Bon at al. 2012) that was obtained by investigating the
long term monitoring spectra of NGC 4151. This investigations continued with a 
serious of papers where many other AGN appeared to show periodicity in their light
curves (Bon et al. 2017, Kova\v cevi\'c et al. 2017, Marziani et al. 2017, etc.).

The main results that we obtained in this investigation are: i) there are indications
in spectral variation (periodicity and line shape varriations) that in some AGN
a SMBH binary can be present, especially in NGC 4151 where spectral variability can be 
explained by SMBH binary dynamics; ii) from modeling of the 
SMBH binaries, taking different parameters of a SMBH binary system, we concluded that
the line shapes and shifts can indicate a SMBH binary, but the dynamical effect of a binary 
system can be hidden by other processes in the BLR.

\section{AGN SPECTRAL CHARACTERISTICS: FROM THE UV TO THE MIR}

As it was noted in the Introduction, basically AGN can be divided in type 1 (with broad lines) and 
type 2 (without broad lines). However, the spectral properties in a wide spectral range from the ultraviolet (UV) to the mid-infrared (MIR) can be quite different in type 1 AGN. As an obvious case is a difference between the narrow line  and broad line Seyfert 1. 

The investigation of the type 1 AGN spectral characteristics is important from two reasons: first is that the correlations between different spectral properties indicate some physical processes (Boroson \& Green 1992); and second is to find some constrains and relationships between spectral characteristics and luminosity, in order to use quasars as standard candles (see e.g. Lusso \& Risaliti 2017).

In our research of AGN spectral characteristics we started from the optical, exploring the relationships
between the broad Balmer lines (mostly H$\beta$ and H$\alpha$) and Fe II features around the H$\beta$ 
(Kova\v cevi\'c et al. 2010, Popovi\'c \& Kova\v cevi\'c 2011) as well between the ratios of Balmer
lines and connection with the continuum (Ili\'c et al. 2012, Rafanelli et al. 2014, Raki\'c et al. 2017). The next step was to connect the spectral characteristics in the optical and UV (Kova\v cevi\'c et al. 2014a, 
Kova\v cevi\'c-Doj\v cinovi\'c \&
Popovi\'c 2015, Joni\'c et al. 2016, Kova\v cevi\'c-Doj\v cinovi\'c et al. 2017) and then to connect the spectral properties between the optical and MIR (Laki\'cevi\'c et al. 2017).

One of very important tasks in this part was to find the stellar population
influence on AGN spectra. With this goal we developed a code for full 
spectrum fitting of AGN spectra (Bon et al. 2014, Bon et al. 2016) using ULYSS code (Koleva 
et al. 2009) - the code for stellar population analysis, that enable us to analyse simultaneously 
complex emission line models, Fe II pseudo continuum, AGN continuum and host galaxy. Using this code we 
investigated properties of Type 2 (see Bon et al. 2014) and
some Type 1 AGN as well (see Bon et al. 2016, Marziani et al. 2017a). 

Additionally, in type 1 AGN we can extract broad line profiles which are important for investigation of the 
BLR structure, especially an accretion disk contribution to broad emission line profiles
(see e.g. Popovi{\'c} et al. 2004, Bon et al. 2006, Bon et al. 2009a, 2009b). We investigated 
the contribution of accretion disk emission using Hamburg-ESO (HE) sample of intermediate to 
high redshift quasars, that are some of the most luminous quasars known, 
hosting very massive black holes. We matched simulated relativistic ray tracing accretion
disk profiles (see Jovanovi\'c 2012), with the optical emission H$\beta$ broad emission
line profiles from HE sample, selected to have centroids of line widths measured at
1/4 of maximum line intensity significantly shifted to the red, in order  to investigate gravitational
redshift in these spectra (see Bon et al. 2015).

The most important obtained results in this part are: i) we constructed a
unique Fe II template in the optical and UV part of AGN spectra (an online program for Fe II fitting 
is available on SerVO site, see
http://servo.aob.rs/FeII$\_$AGN/);
ii) we proposed a model for Balmer quasi-continuum in the UV part of AGN spectra and also find that 
the nature of the optical Fe II lines is different than the UV Fe II ones; 
iii) using polycyclic aromatic hydrocarbons (PAH) in the MIR we found that the division between type 1 and type 2 AGN using optical criteria does not follow the MIR characteristics; iv) we developed a code for fitting 
the stellar population simultaneously with the spectrum of an AGN (type 1 and type 2) that can
be used in investigation of the characteristics of AGN emission and host stellar population.

\section{EXPLORING AGN STRUCTURE: MODELING  AND GRAVITATIONAL LENSING}

One of the important way to explore the structure of AGN is by modeling their emission regions. In this part 
we developed a model of relativistic accretion disk around SMBH, 
and simulated its X-ray radiation in the Fe K$\alpha$ line (see 
Jovanovi\'{c} 2012, and references therein). Comparisons between the 
simulated and observed Fe K$\alpha$ line profiles were then used to 
determine the parameters of the relativistic accretion disk, such as 
inclination, emissivity, inner and outer radius, as well as the spin of 
the central SMBH (see e.g. Jovanovi\'{c} et al. 2011, 2016).

\begin{figure}[ht!]
\centering
\includegraphics[width=0.56\textwidth]{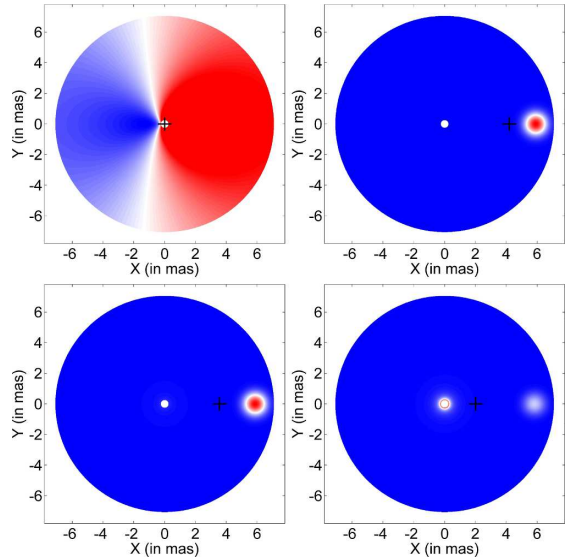}
\caption{Photocenter positions (crosses) in the case of an accretion 
disk without (top left) and with perturbation (next three panels) for
three different values of emissivity index (Popovi\'{c} et al. 2012).}
\label{fig01}
\end{figure}

The accretion disk model can be applied to the optical line emission, and comparing
the observations and disk-models we demonstrated that 
the variability in broad line shapes can be explained by the hot-spot motion in the disk (Jovanovi\'c et al. 2010).

On the other hand, we also constructed a model for the dusty torus that was developed
within SKIRT code, a state-of-the-art radiative transfer 
code based on Monte Carlo technique (Stalevski et al. 2011, Stalevski et al. 2012a),
and explore the emission of the dusty torus with different physical parameters. 
More recently, dust emission models were used to study the relation
between the ratio of the torus and AGN luminosities and the dust
covering factor. This study (Stalevski et al. 2016) found that the
observed luminosity ratio very often under- or overestimates the actual
covering factor and provided a novel way to correct it and obtain the
true values. In another recent study, the detailed modeling of the dust
emission of the archetypal type 2 AGN in Circinus galaxy was performed,
showing that contrary to the expectations, a major part of the dust
emission is coming from the polar region, in a form of cone-like
outflows (Stalevski et al. 2017).

The models of accretion disk (which emits in X-ray and optical) and 
dusty torus are used to explore some variability that can be seen in quasars,
which is important for the Gaia reference frame (Popovi\'c et al. 2012). As an example of the
photo-center variability  we illustrate in Fig.  \ref{fig01} the displacement of disk's photo-center 
as a function of the bright spot emissivity.

The  AGN central part  model (disk+torus, described above) allowed us to explore gravitational 
lensing effects on the spectra of lensed quasars. Additionally we modeled milli- and micro-lens maps, and simulate the lens transition
across the inner part of an AGN, modeling the spectral variations (Simi\'c et al. 2011, Stalevski et al. 2012b, 
Popovi\'c \& Simi\'c 2013, Simi\'c \& Popovi\'c 2014). Also we performed observations of several lensed 
quasars with 6m SAO telescope in order to compare our models with observations (Popovi\'c et al. 2010)

In this part we can outline the following results: i) we developed a unique model for the AGN torus
(note here that paper of Stalevski et al. 2012 has been cited more than 100 times); 
and a library of emission models of the AGN dusty torus is available online 
(https://sites.google.com/site/skirtorus/); ii) we give prediction for variation in 
the quasar position using the models from accretion disk to torus that is
very useful for the Gaia reference frame; iii) we give predictions in spectral 
variation due to microlensing that can 
be used in the separation between the intrinsic quasar variation (see Section 2) and a microlensing event.

\section{BLACK HOLE MASSES - MEASUREMENT AND VIRIALIZATION}

\begin{wrapfigure}{r}{0.45\textwidth}
  \begin{center}
    \includegraphics[width=0.4\textwidth]{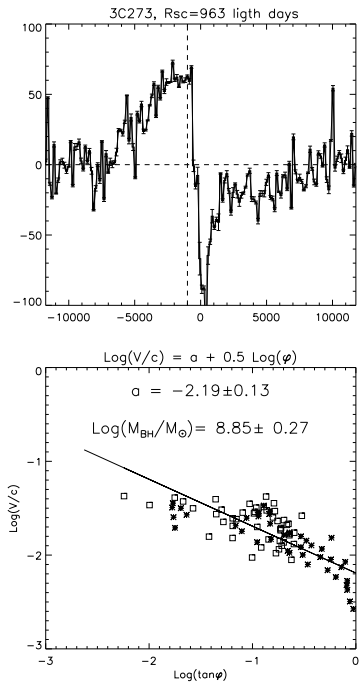}
  \end{center}
  \caption{The polarization angle - PA observed in quasar 3C 273 (upper panel), 
  and estimation of the black hole mass of quasar (bottom panel) as given in Afanasiev \& Popovi\'c (2015). }
\end{wrapfigure}
As noted in the Introduction, it seems that the central SMBH has influence on the structure of
host galaxy and its evolution. Therefore the measurement of SMBH masses in center of galaxies is a
very important task in astronomy today.
In difference with "normal" galaxies, galaxies with AGN in the center give us possibility to measure 
the mass of central black hole exploring the gas motion in the BLR. There are several direct and indirect 
methods for 
the black hole measurement (see review of Peterson 2014), among them the reverberation is one of the 
direct methods, but it is also telescope time consuming. 

In this part, we worked in two directions: i) exploring the geometry and structure of the BLR using 
variability (see references in Section 2) and comparing virialization in different broad
lines (Mg II and H$\beta$, see Joni\'c et al. 2016), and ii) exploring polarization in
broad lines as a tool to measure the SMBH mass (Afanasiev et al. 2014, Afanasiev \& Popovi\'c 2015, Savi\'c et al. 2017).

To measure SMBH masses using polarization in the emission line we 
performed observations of a number of AGN with 6-m SAO telescope and we explore
observed data to find black hole masses. Additionally, we did simulations using STOKES code 
(Goosmann et al. 2013) and find that the proposed method can give very good results (Savi\'c et al. 2017).

%\begin{figure}[ht!]
%\centering
%\includegraphics[width=0.46\textwidth]{Angle_VS_Vel.ps}
%\caption{}
%\label{fig02}
%\end{figure}

The important results we obtained in this part is that we give a new method for
AGN black hole mass measurements
using the polarization in broad lines (illustrated in Fig. 3). The method can
explore virialization in the BLR, and can be applied on the one-epoch observations
(see Afanasiev \& Popovi\'c 2015 for more details).

\section{AGN GAMMA RAY EMISSION, GAMMA RAY BURSTS AND THEIR IMPACT ON THE LOW IONOSPHERE}

The gamma ray radiation can be observed in objects with jets, as e.g. blazars, but also can be produced by
exotic objects (as e.g. black hole collisions) which emit enormous gamma ray flux, known as gamma ray bursts (GRBs). Gamma ray emission is mostly
connected with violent processes in the Universe, and has been the subject of investigation for the last
several decades (after the lunching of gamma ray telescopes). 

In this part we investigate the gamma-ray emission in  blazars. First we explore the extraordinary gamma-ray 
activity of the gravitationally lensed blazar PKS 1830-211 (Donnarumma et al. 2011), finding also that this 
variability can be caused by microlensing (since the source of gamma radiation is very compact),
that was a direct application of our AGN and lens models (see Section 5). Second, we explore flare-like variability of the Mg II  emission line in the
gamma-ray blazar 3C 454.3 (Le\'on-Tavares et al. 2013), 
connecting the Mg II emission variability and variability in gamma ray emission. It was 
interesting that we found 
a good correlation between gamma-ray and Mg II variability, indicating that Mg II originates from the jet.

To explore the origin of the gamma-rays in shock-waves, we developed a shock wave model 
and fit 30 GRBs (Simi\'c \& Popovi\'c 2012).
We found some characteristics of GRBs, and divide them in two groups - short and long lasting GRBs.

A GRB emits the huge amount of energy that impacts the upper parts of Earth atmosphere, and 
this opens a question: how much
a GRB can affect the Earth atmosphere, especially the low ionosphere? This was a subject of our 
research (Nina et al. 2015), and we found that  GRBs perturb the low ionosphere, and its
reaction is significant in a short period of several seconds after the GRB has been detected by satellites.

The most important results in these investigations are: i) we found that gamma ray emission in some 
AGN correlates with the broad Mg II line, this indicates that in some cases broad Mg II line is originated
in the jet-like region, 
that it is not connected with the classical BLR -- this should be taken into account when Mg II line is used for 
black hole mass measurements; ii) we found statistical significance that  GRBs 
have influence on the ionosphere, since we found that reaction of ionosphere is several 
seconds after the GRB detection by satellites (this discovery was noted as 'research spotlight' in March, 2016,
see https://eos.org/research-spotlights/gamma-ray-bursts-leave-their-mark-in-the-low-ionosphere).

\section{SUMMARY AND FUTURE PLANS}

Here we present the most important results of the 
spectro-polarimetric observations and modeling of different parts of AGN (accretion disk, BLR, dusty torus),
obtained in the last several years. We also note, that beside the scientific part, the project was a base for worldwide collaboration and  some research subjects were PhD and master thesis for several students. Additionally, we organized several workshops (see http://servo.aob.rs/eeditions/Workshops.php) and spectral line shapes conferences (see http://servo.aob.rs/scslsa11/).

Probably the project will finish at the end of this year or during the next one, but we are going to continue our activities in the field of spectroscopy of extragalactic objects. We are going to: i) continue with the monitoring of several broad line AGN in order to find the geometry and size of the BLRs and estimate SMBH masses; ii) continue with spectropolarimetric observations of several type 1 AGN in order to measure masses and find the inclination of the BLR; iii) develop models of AGN central part and 
gravitational lenses in order to explore the influence of milli- and micro-lensing to the spectra of lensed quasars; iv) explore spectral properties of type 1 AGN, including also X-ray emission, in order to find some relationships which can be used for cosmological investigations.

At the end let us note that we are included in the Large Synoptic Survey Telescope - LSST project (in scientific part for AGN investigation),  that will give us opportunity to extend our investigations. Also, we are going to provide low resolution spectrograph which can be installed at 1.4m telescope Milankovi\'c, and can be actively involved in AGN monitoring campaign.

%\vfil\newpage

% Table
%\begin{table}
%\begin{center}
%\begin{tabular}{|c|c c|}
%\hline
%$t$    &$\theta (^o)$ & $\rho ('')$ \\
%\hline
%2002.0 &    174.0     &    0.259    \\
%2003.0 &    174.7     &    0.262    \\
%2004.0 &    175.3     &    0.265    \\
%2005.0 &    175.9     &    0.268    \\
%2006.0 &    176.6     &    0.270    \\
%\hline
%\end{tabular}
%\caption{Calculated positions for the next five years}
%\end{center}
%\end{table}

% Figure (in PS or EPS format)
%\begin{figure}
%\includegraphics[width=12cm,height=8cm]{figure.ps}
%\caption{?????}
%\end{figure}

% Equation
%\begin{equation}
%G(x)=\int_0^\infty e^{-{{x^2}\over{2}}} dx
%\end{equation}

% References
\references
 
Afanasiev, V. L., Popovi\'c, L. \v C.: 2015, \journal{ApJ}, \vol{800L}, 35
 
Afanasiev, V. L., Popovi\'c, L. \v C., Shapovalova, A. I., Borisov, N. V., Ili\'c, D.: 2014, \journal{MNRAS}, \vol{440}, 519

Afanasiev, V. L., Shapovalova, A. I.,  Popovi\'c, L. \v C., Borisov, N. V.: 2015, \journal{MNRAS}, \vol{448}, 2879

Begelman, M. C., Blandford, R. D., Rees, M. J.: 1980, \journal{Nature}, \vol{287}, 307

Bon, N., Bon, E., Marziani, P.,  Jovanovi{\'c}, P.: 2015, \journal{APSS}, \vol{360}, 7 

Bon, E., Gavrilovi{\'c}, N., La Mura, G., Popovi{\'c}, L.~{\v C}.: 2009a, \journal{NewAR}, \vol{53}, 121

Bon, E., Jovanovi{\'c}, P., Marziani, P., et al.: 2012, \journal{ApJ}, \vol{759}, 118

Bon, E., Marziani, P.,  Bon, N.: 2017,  \journal{New Frontiers in Black Hole Astrophysics}, \vol{324}, 176 

Bon, N., Popovi\'c, L. \v C., Bon, E.: 2014,\journal{AdSpR}, \vol{54}, 1389

Bon, E., Popovi{\'c}, L.~{\v C}., Gavrilovi{\'c}, N., La Mura, G.,
 Mediavilla, E.: 2009b, \journal{MNRAS}, \vol{400}, 924

Bon, E., Popovi{\'c}, L.~{\v C}., Ili{\'c}, D., Mediavilla, E.: 2006, \journal{NewAR}, \vol{50}, 716

Bon, E., Zucker, S., Netzer, H., et al.: 2016, \journal{ApJS}, \vol{225}, 29 

Boroson, T. A., Green, R. F.: 1992, \journal{ApJS}, \vol{80}, 109

Donnarumma, I., De Rosa, A., Vittorini, V., Miller, H. R., Popovi\'c, L. \v C., Simi\'c, S. et al.: 2011, \journal{ApJ}, \vol{736L}, 30

Gaskell, C. M.:  1983, \journal{Liege International Astrophys. Colloq.}, \vol{24}, 473

Goosmann, R.~W., Gaskell, C.~M., \& Marin, F.: 2014,  \journal{AdSpR}, \vol{54}, 1341
	
Ili\'c, D., Popovi\'c, L. \v C., La Mura, G., Ciroi, S., Rafanelli, P.: 2012, \journal{A\&A}, \vol{543A}, 142

Ili{\'c}, D., Popovi{\'c}, L.~{\v C}., Shapovalova, A.~I., et al.: 2015, \journal{JApA}, \vol{36}, 433 

Ili{\'c}, D., Shapovalova, A.~I., Popovi{\'c}, L.~{\v C}., et al.: 2017, \journal{FrASS}, \vol{4}, 12 

Joni\'c, S., Kova\v cevi\'c-Doj\v cinovi\'c, J., Ili\'c, D., and Popovi\'c, L. \v C.: 2016, \journal{Ap\&SS}, \vol{361}, 101

Jovanovi\'{c}, P.: 2012, \journal{NewAR}, \vol{56}, 37.

Jovanovi\'{c}, P., Borka Jovanovi\'{c}, V., Borka, D.: 2011, \journal{Baltic Astronomy}, \vol{20}, 468.

Jovanovi\'{c}, P., Borka Jovanovi\'{c}, V., Borka, D., 
Bogdanovi\'{c}, T.: 2014, \journal{AdSpR}, \vol{54}, 1448.

Jovanovi\'c, P., Borka Jovanovi\'c, V.,  Borka, D.,  Popovi\'c, L. \v C.: 2016,\journal{Ap\&SS}, \vol{361}, 75

Jovanovi{\'c}, P., Popovi{\'c}, L.~{\v C}., Stalevski, M., \& Shapovalova, A.~I.: 2010, \journal{ApJ}, \vol{718}, 168 

Koleva, M., Prugniel, P., Bouchard, A., Wu, Y.: 2009, \journal{A\&A}, \vol{501}, 1269

Kova\v cevi\'c, J., Popovi\'c L. \v C., Dimitrijevi\'c, M.S.: 2010, \journal{ApJS}, \vol{189}, 15

Kova\v cevi\'c, J., Popovi\'c, L. \v C., Kollatschny, W.: 2014a, \journal{AdSpR}, \vol{54}, 1347

Kova{\v c}evi{\'c}, A., Popovi{\'c}, L.~{\v C}., Shapovalova, A.~I., et 
al.: 2014,  \journal{AdSpR}, \vol{54}, 1414

Kova\v cevi\'c, A., Popovi\'c, L. \v C., Shapovalova, A. I., Ili\'c, D., Burenkov, A. N., Chavushyan, V. H.: 2015, \journal{JApA}, \vol{36}, 475

Kova{\v c}evi{\'c}, A., Popovi{\'c}, L.~{\v C}., Shapovalova, A.~I., \& Ili{\'c}, D.: 2017, \journal{Ap\&SS}, \vol{362}, 31 

Kova\v cevi\'c-Doj\v cinovi\'c, Mar\v ceta-Mandi\'c, S., Popovi\'c, L. \v C.: 2017, \journal{FrASS}, \vol{4}, 7

Kova\v cevi\'c-Doj\v cinovi\'c, J., Popovi\'c, L. \v C.: 2015, \journal{ApJS}, \vol{189}, 15

Laki\'cevi\'c, M., Kova\v cevi\'c-Doj\v cinovi\'c, J., Popovi\'c, L. \v C.: 2017, \journal{MNRAS}, \vol{472}, 334

Le\'on-Tavares, J., Chavushyan, V., Pati\~no-Álvarez, V., Valtaoja, E., Arshakian, T. G., Popovi\'c, L. \v C.. et al.: 2013, \journal{ApJ}, \vol{763L}, 36

Lusso, E., Risaliti, G.: 2017, \journal{A\&A}, \vol{602A}, 79

Marziani, P., Bon, E., Grieco, A., et al.: 2017, \journal{New Frontiers in Black Hole Astrophysics}, \vol{324}, 243 

Marziani, P., Negrete, C.~A., Dultzin, D., et al.: 2017a, \journal{New Frontiers in Black Hole Astrophysics}, \vol{324}, 245 

Nina, A., Simi\'c, S.,  Sre\'ckovi\'c, V. A., Popovi\'c, L. \v C.: 2015, \journal{GeoRL}, \vol{42}, 8250	

Peterson, B. M.: 2014, \journal{SSRv}, \vol{183}, 253
	
Popovi\'c, L. \v C.: 2012, \journal{NewAR}, \vol{56}, 74
 
Popovi\'c, L. \v C., Jovanovi\'c, P., Stalevski, M., Anton, S., Andrei, A. H., Kova\v evi\'c, J., Baes, M.: 2012, \journal{A\&A}, \vol{538A}, 107

Popovi\'c, L. \v C.,  Kova\v cevi\'c, J.: 2011, \journal{ApJ}, \vol{738}, 68
 
Popovi\'c, L. \v C., Mediavilla, E., Bon, E., \& Ili\'c, D.: 2004, \journal{A\&A}, \vol{423}, 909

Popovi\'c, L. \v C., Mediavilla, E. G., Pavlovi\'c, R.: 2000, \journal{SerAJ}, \vol{162}, 1
 
Popovi\'c, L. \v C., Moiseev, A. V., Mediavilla, E.,  Jovanovi\'c, P., Ili\'c, D.,  Kova\v evi\'c, J.,
Mu\~noz, J. A.: 2010, \journal{ApJ}, \vol{721}, 139

Popovi\'c, L. \v C., Shapovalova, A. I., Chavushyan, V. H., Ili\'c, D.,  Burenkov, A. N., Mercado, A.  Bochkarev, N. G.: 2008, \journal{PASJ}, \vol{60}, 1

Popovi{\'c}, L.~{\v C}., Shapovalova, A.~I., Ili{\'c}, D., et al.: 2011, \journal{A\&A}, \vol{528}, 130

Popovi{\'c}, L.~{\v C}., Shapovalova, A.~I., Ili{\'c}, D., et al.: 2014, \journal{A\&A}, \vol{572}, A66 

Popovi\'c, L. \v C.; Simi\'c, S.: 2013, \journal{MNRAS}, \vol{432}, 848

Rafanelli, P., Ciroi, S., Cracco, V., Di Mille, F., Ili\'c, D., La Mura, G., Popović, L. \v C.: 2014, \journal{AdSpR}, \vol{54}, 1362
	
Raki{\'c}, N., La Mura, G., Ili{\'c}, D., Shapovalova, A. I., Kollatschny, W., Rafanelli, P.,
Popovi\'c, L. \v C.: 2017, \journal{A\&A}, \vol{603}, A49 

Shapovalova, A.I., Burenkov, A. N., Carrasco, et al.: 2001, \journal{A\&A}, \vol{376}, 775

Shapovalova, A. I., Doroshenko, V. T., Bochkarev, N. G. et al.: 2004, \journal{A\&A}, \vol{422}, 925

Shapovalova, A. I., Popovi\'c, L. \v C., Collin, S., et al. : 2008, \journal{A\&A}, \vol{486}, 99S

Shapovalova, A.~I., Popovi{\'c}, L.~{\v C}., Burenkov, A.~N., et al.: 2010a, \journal{A\&A}, \vol{517}, A42 

Shapovalova, A.~I., Popovi{\'c}, L.~{\v C}., Burenkov, A.~N., et al.: 2010b, \journal{A\&A}, \vol{509}, A106 

Shapovalova, A.~I., Popovi{\'c}, L.~{\v C}., Burenkov, A.~N., et al.: 2012, \journal{ApJS}, \vol{202}, 10 

Shapovalova, A.~I., Popovi{\'c}, L.~{\v C}., Burenkov, A.~N., et al.: 2013, \journal{A\&A}, \vol{559}, A10

Shapovalova, A.~I., Popovi{\'c}, L.~{\v C}., Chavushyan, V.~H., et al.: 2016, \journal{ApJS}, \vol{222}, 25 

Shapovalova, A.~I., Popovi{\'c}, L.~{\v C}., Chavushyan, V.~H., et al.:  2017, \journal{MNRAS}, \vol{466}, 4759 

Savi\'c, Dj., Goosmann, R., Popovi\'c, L. \v C., Marin, F., Afanasiev, V. L.: 2017, sent to A\&A

Simi\'c, S., Popovi\'c, L. \v C.: 2014, \journal{AdSpR}, \vol{54}, 1439

Simi\'c, S., Popovi\'c, L. \v C.: 2016, \journal{Ap\&SS}, \vol{361}, 59

Simi\'c, S., Popovi\'c, L. \v C.: 2012, \journal{IJMPD}, \vol{21}, 1250028

Simi\'c, S., Popovi\'c, L. \v C., Jovanovi\'c, P.: 2011, \journal{BaltA}, \vol{20}, 481

Smailagi{\'c}, M., Bon, E.: 2015,  \journal{JApA}, \vol{36}, 513 
 
Stalevski, M., Fritz, J., Baes, M., Nakos, T., Popovi\'c, L. \v C.: 2011, \journal{BaltA}, \vol{20}, 490

Stalevski, M., Fritz, J., Baes, M., Nakos, T.,  Popovi\'c, L. \v C.: 2012a,  \journal{MNRAS}, \vol{420}, 2756

Stalevski, M., Jovanovi\'c, P., Popovi\'c, L. \v C., Baes, M.: 2012b, \journal{MNRAS}, \vol{425}, 1576
 
Stalevski, M., Ricci, C., Ueda, Y., et al.: 2016, \journal{MNRAS}, \vol{458}, 2288

Stalevski, M., Asmus, D., \& Tristram, K.~R.~W.: 2017, \journal{MNRAS}, \vol{472}, 3854
 
Woo, J. H., Cho, H., Husemann, B., Komossa, S., Park, D. \& Bennert, V. N.: 2014, \journal{MNRAS}, \vol{437}, 32

\endreferences

\end{document}